# Long memory and multifractality: A joint test

John Goddard

Bangor Business School, Bangor University, Gwynedd, UK

Enrico Onali

Aston Business School, Aston University, Birmingham, UK

**Abstract**

The properties of statistical tests for hypotheses concerning the parameters of the multifractal model of asset returns (MMAR) are investigated, using Monte Carlo techniques. We show that, in the presence of multifractality, conventional tests of long memory tend to over-reject the null hypothesis of no long memory. Our test addresses this issue by jointly estimating long memory and multifractality. The estimation and test procedures are applied to exchange rate data for 12 currencies. In 11 cases, the exchange rate returns are accurately described by compounding a NIID series with a multifractal time-deformation process. There is no evidence of long memory.

# Long memory and multifractality: A joint test

1.  **Introduction**

The statistical properties of financial asset returns are of key importance for finance theory, and for portfolio and risk management. The econophysics literature has grown exponentially over the last couple of decades, and fractal models have been employed to explain empirical regularities at odds with mainstream financial economics theory, such as: power-laws and self-similarity [1], long memory in returns and volatility [2-4], and multifractality [5]. However, so far there is no study that addresses the problem of estimating multifractality in the presence of long memory, and whether long memory can affect estimation of multifractality, and vice versa.

The multifractal model of asset returns (MMAR) is capable of accommodating exceptional events (large shocks), and can represent either Normal or non-Normal log returns with a finite variance [6]. The MMAR nests the fractionally-integrated ARFIMA(0,d,0) model, which allows for long memory in returns, and can also accommodate long memory in volatility associated with multifractality in the trading process. Long memory in returns does not necessarily imply multifractality, and vice versa. Thus, the MMAR is able to replicate the pricing behaviour of many types of financial assets.

The MMAR describes a continuous-time process, constructed by compounding fractional brownian motion (FBM) with a random, multifractal time-deformation process. The time-deformation process allows for volatility clustering in log returns measured at any scale, and for long memory in volatility. The property of multifractality is identified empirically through estimation of the scaling function, $E(|\Delta^{(n)}p_t|^q)$, where $p_t$ is log price, and $\Delta^{(n)}p_t = p_t - p_{t-n}$ is the log return measured over the scale n, over a range of values for q.

Previous empirical studies employ methods such as Multi-Fractal Detrended Fluctuation Analysis [5, 7, 8, 9, 10], Wavelet Transform Modulus Maxima [7] or a Generalized Hurst Exponent approach [11] to detect multifractality. Multifractality has been interpreted as an indicator of market inefficiency [12], and it has been suggested that multifractality might be generated by herding behaviour [13]. However, other studies suggest that indications of multifractality might simply reflect fat-tailed returns [14]. These inconclusive results suggest that the methods employed to detect multifractality are far from perfect, and are in need of refinement.

Despite the ability of the MMAR to replicate stylized facts for financial asset returns (such as long memory in volatility, multi-scaling, and fat-tails), the number of empirical studies based on the MMAR is rather small [6,15]. We believe this dearth of empirical evidence is, at least in part, a consequence of difficulties in the interpretation of point estimates of the MMAR parameters. Furthermore, most previous studies fail to consider the joint estimation of long memory and multifractality parameters.

In this paper, we provide statistical testing criteria for estimated long memory and multifractality parameters, based on conventional hypothesis testing methodology. In so doing, we facilitate comparisons between processes that are described by the MMAR, and processes that are characterized by models nested within the MMAR. The latter include, for example, NIID (normal, independent and identically distributed) returns, and long-range dependent returns generated from a log price series characterized by FBM. Monte Carlo simulations are used to generate critical values for the relevant tests. The methods are illustrated by means of an analysis of the long memory and multifractal properties of the daily log returns for the exchange rates of 12 currencies against the US dollar for the period 1993-2012.

## 2. The multifractal model of asset returns

According to the MMAR, $p_t = B_H[\theta(t)]$, where $B_H[\ ]$ denotes FBM, and $\theta(t)$ denotes a time-deformation process. In order to construct $\theta(t)$, consider first the case $T=2^K$ for some integer value of K. The specification of $\theta(t)$ is

$$\Delta\theta(t) = \theta(t)-\theta(t-1) = T\Omega^{-1}\prod_{k=1}^{K} m[\eta_t(k)], \text{ where } \Omega = \sum_{\tau=1}^{T}\prod_{k=1}^{K} m[\eta_\tau(k)] \quad (1)$$

where $\eta_t(k) = h$ if $2^{-k}(h-1)T+1 \leq t \leq 2^{-k}hT$ for $k=1...K$, $h=1...k$, and $t=1...T$; and the multiplier $m[\eta_t(k)]$ is assumed to be drawn randomly from a lognormal distribution with mean $\lambda$ and variance $\sigma^2 = 2(\lambda-1)/\ln 2$ for $\lambda>1$.

In the case $T \neq 2^K$ for any integer value of K, the following adjustments are required. Let $K^*$ denote the minimum value of K such that $T<2^K$. The multipliers $m[\eta_s(k)]$ are constructed in accordance with the procedure described above, for $k=1...K^*$ and $s=1...2^{K^*}$.

$$\Delta\theta(t) = T\Omega^{-1}\prod_{k=1}^{K} m[\eta_{r+t}(k)], \text{ where } \Omega = \sum_{\tau=r+1}^{r+T}\prod_{k=1}^{K} m[\eta_{r+\tau}(k)] \quad (2)$$

where r is a randomly drawn integer, distributed uniformly over the interval $(0, K^*-T)$.

Let $u_t \sim N(0,\Delta\theta(t))$. For $\lambda=1$, $\Delta\theta(t)=1$ for all t. $u_t$ is homoscedastic and there is no multifractality in $u_t$. For $\lambda>1$, $u_t$ is heteroscedastic and there is multifractality in $u_t$. Combining the multifractal time-deformation process with FBM such that $p_t = B_H[\theta(t)]$, the data generating process for $\Delta^{(1)}p_t$ is

$$(1-L)^d \Delta^{(1)}p_t = u_t \quad (3)$$

where L denotes the lag operator $L^s\Delta^{(1)}p_t = \Delta^{(1)}p_{t-s}$ for $s=0,1,2,...$

Using the Wold decomposition, the moving average representation of FBM is

$$\Delta^{(1)}p_t = (1-L)^{-d}u_t = [1 + dL + \{d(d-1)/2!\}L^2 + \{d(d-1)(d-2)/3!\}L^3 + ...]u_t \quad (4)$$

$\Delta^{(1)}p_t$ is stationary for $-0.5<d<0.5$. Let $H=d+0.5$ denote the Hurst exponent. The variance of log returns has the following scaling property

$$\text{var}[\Delta^{(n)}p_t] = n^{2H}\text{var}[\Delta^{(1)}p_t] \qquad (5)$$

Let $\gamma_k = E(\Delta^{(1)}p_t\Delta^{(1)}p_{t-k})/E(\Delta^{(1)}p_t^2)$ for $k=0,1,2,...$ denote the autocovariance function of $\Delta^{(1)}p_t$. For $d=0$ ($H=0.5$), $p_t$ is a martingale, and $\gamma_k=0$ for $k\geq 1$. For $-0.5<d<0$ ($0<H<0.5$), $\Delta p_t$ exhibits negative persistence, and $\gamma_k<0$ for $k\geq 1$. For $0<d<0.5$ ($0.5<H<1$), $\Delta p_t$ exhibits positive persistence and long memory, and $\gamma_k>0$ for $k\geq 1$.

3.   **Estimation of the parameters of the MMAR**

Following the method described by [6], estimation of the two parameters of interest in the MMAR, $\{H, \lambda\}$, proceeds as follows. Starting from the first observation, subdivide the sample period T into M contiguous subperiods, each containing n observations such that $T-n < Mn \leq T$. Let $v_m$ denote the absolute value of the log return calculated over the n observations within subperiod m, for $m=1...M$, $v_m = |p_{mn} - p_{m(n-1)}|$.

If $Mn<T$, then $L=T-nM$ observations at the end of the sample period are unused in the calculation of $\{v_m\}$. In order to incorporate these observations into the analysis, the above calculation is repeated, starting from the L+1th observation. A second set of M values of $\{v_m\}$ is obtained, labelled (for convenience) $\{v_{m+1}...v_{2M}\}$. If $Mn=T$, $\{v_1...v_M\}$ and $\{v_{M+1}...v_{2M}\}$ are identical. The q'th-order partition function for scale n is

$$S_q(T,n) = 2^{-1}\sum_{m=1}^{2M}(v_m)^q \qquad (6)$$

$S_q(T,n)$ is calculated for various values of q, and for each q for various values of n. The scaling behaviour of $S_q(T,n)$ is investigated by examining the power law relationship

$$E[S_q(T,n)] = Tc(q)\times n^{qh(q)} = Tc(q)\times n^{\tau(q)+1} \qquad (7)$$

where c(q) is the prefactor, h(q) is the generalized Hurst exponent, and $\tau(q) = qh(q)-1$ is the scaling function.

In the case of unifractality, the scaling behaviour of $S_q(T,n)$ is the same for all q, or h(q)=H for all q. In the case of multifractality, the scaling behaviour of $S_q(T,n)$ varies with q, or h(q) depends upon q. Let $q^*$ denote the value of q such that $q^*h(q^*)=1$ and $\tau(q^*)=0$. This definition implies

$$H=h(q^*) \quad \text{or} \quad H=1/q^* \tag{8}$$

If the data generating process for returns is the MMAR with lognormal multipliers and $\lambda>1$, the scaling function $\tau(q)$ is quadratic in q

$$\tau(q) = \tau_0+\tau_1 q+\tau_2 q^2 \tag{9}$$

For convenience and without any loss of generality, the intercept of (9) may be normalized, $\tau_0 = -1$. Using (7) and (9), $\tau(q)$ can be estimated from the fixed-effects regression

$$\ln[S_q(T,n)] = a(q) + [-1+\tau_1 q+\tau_2 q^2] \ln(n) + \text{error} \tag{10}$$

where $a(q) = \ln[nTc(q)]$.

In the previous literature, evidence of multifractality is commonly obtained from inspection of the multifractal spectrum, $f(\alpha) = \min_q [\alpha q - \tau(q)]$. Let $\alpha_0$ denote the value of $\alpha$ that maximizes $f(\alpha)$. Using (8), it is easily shown

$$\alpha_0=\tau_1 \quad \text{and} \quad H=2\tau_2[(\tau_1^2+4\tau_2)^{1/2}-\tau_1]^{-1} \tag{11}$$

Equation (11) provides an estimation method for H. A test of $H_0$:H=0.5 against $H_1$:H≠0.5 is a test for NIID returns under $H_0$, against either positive persistence and long memory (H>0.5), or negative persistence (H<0.5) under $H_1$.

Figure 1 illustrates the interpretation of the multifractal spectrum. $f(\alpha)$ can be interpreted as the lower envelope of the set of linear functions $\alpha q - \tau(q)$. In addition to $\alpha_0$, other reference points in Figure 1 are $(\alpha_1, \alpha_1/H)$, $(\alpha_{MIN}, 0)$, and $(\alpha_{MAX}, 0)$. Since $\tau(q^*)=0$ and $q^*=1/H$, the linear function $\alpha q^* - \tau(q^*)$ intersects the origin, and is tangential to $f(\alpha)$ at the point $(\alpha_1, \alpha_1/H)$. $\alpha_{MIN}$ and $\alpha_{MAX}$ are the minimum and maximum values of $\alpha$ for which $f(\alpha) \geq 0$.

In the case of unifractality, the scaling function is $\tau(q)=-1+Hq$, and all of the lines of tangency pass through the same point, $(H, 1)$. The lower envelope is degenerate, and the multifractal spectrum has the appearance of a spike, located at $\alpha_0=\alpha_1=\alpha_{MIN}=\alpha_{MAX}=H$, such that $f(\alpha)=1$ for $\alpha=H$, and $f(\alpha)=-\infty$ for $\alpha \neq H$.

The parameter $\lambda$ is estimated using the relationship

$$\alpha_0 = \lambda H \qquad \text{or} \qquad \lambda = \alpha_0/H \tag{12}$$

In the case of unifractality, $\alpha_0=H$ implies $\lambda=1$. Accordingly, test of $H_0:\lambda=1$ against $H_1:\lambda>1$ is a test for unifractality (under $H_0$) against multifractality (under $H_1$).

**4.  Hypothesis tests for the presence of long memory and multifractality**

The hypothesis tests reported in this study are derived from the empirical distributions of the estimators of H and $\lambda$ based on (11) and (12), obtained from 5,000 replications of NIID series and denoted $\{\overline{H}, \overline{\lambda}\}$. We examine tests of the following hypotheses: (i) $H_0:H=0.5$ against $H_1:H \neq 0.5$; (ii) $H_0:\lambda=1$ against $H_1:\lambda>1$; and (iii) $H_0:\{\lambda=1, H=0.5\}$. Using $\{\overline{H}, \overline{\lambda}\}$ obtained from the Monte Carlo simulation and the estimated H and $\lambda$ for an observed series, denoted $\{\hat{H}, \hat{\lambda}\}$, the p-value for (i) is $2\min\{\pi_H, 1-\pi_H\}$ where $\pi_H = 5000^{-1}\sum 1_{\overline{H}>\hat{H}}$, $1_{\overline{H}>\hat{H}}$ is the indicator function for $\overline{H} > \hat{H}$, and the summation is over the 5,000 simulated $\overline{H}$. The p-value for (ii) is $\pi_\lambda = 5000^{-1}\sum 1_{\overline{\lambda}>\hat{\lambda}}$, where $1_{\overline{\lambda}>\hat{\lambda}}$ is the indicator function for $\overline{\lambda} > \hat{\lambda}$ and the summation is the same. The p-value for (iii), denoted $\pi_{H,\lambda}$, is obtained iteratively as follows:

1. On the first iteration (j=1), set the initial p-value to p=1–0.001j.

2. Using the simulated $\{\overline{H}, \overline{\lambda}\}$, fit the confidence ellipse

$$\gamma_1\lambda^2 + \gamma_2 H^2 + \gamma_3\lambda H + \gamma_4\lambda + \gamma_5 H = \gamma_0(p)$$

3. For a significance level of α=p, $H_0$ is rejected under any of the following conditions:

$$\hat{H} < H_{MIN}\ ;\quad \hat{H} > H_{MAX}\ ;\quad \{\lambda_1 \leq \hat{\lambda} \leq \lambda_{MAX}\ and\ \hat{H} < f_1(\hat{\lambda})\}\ ;$$

$$\{\lambda_2 \leq \hat{\lambda} \leq \lambda_{MAX}\ and\ \hat{H} > f_2(\hat{\lambda})\}$$

where $f_1(\lambda) = 2^{-1}\{-(\gamma_3\lambda + \gamma_4) - [(\gamma_3\lambda + \gamma_5)^2 - 4\gamma_2(\lambda^2 + \gamma_4\lambda - \gamma_0(p))]^{1/2}\}$

$f_2(\lambda) = 2^{-1}\{-(\gamma_3\lambda + \gamma_4) + [(\gamma_3\lambda + \gamma_5)^2 - 4\gamma_2(\lambda^2 + \gamma_4\lambda - \gamma_0(p))]^{1/2}\}$

$\lambda_1 = \arg\min_\lambda f_1(\lambda)$; $H_{MIN} = f_1(\lambda_1)$; $\lambda_2 = \arg\max_\lambda f_2(\lambda)$; $H_{MAX} = f_2(\lambda_2)$

$g(H) = 2^{-1}\{-(\gamma_3 H + \gamma_4) + [(\gamma_3 H + \gamma_4)^2 - 4(\gamma_2 H^2 + \gamma_5 H - \gamma_0(p))]^{1/2}\}$

$H_2 = \arg\max_H g(H)$; $\lambda_{MAX} = g(H_2)$

4. If $H_0$ is rejected at step 3, proceed to the second iteration by resetting j=2 and p=1–0.001j, and repeat steps 2,3. The procedure is repeated for further iterations (j=3,...,999), and terminates when a value for p is obtained at which $H_0$ is not rejected.

Figure 2 illustrates the construction of the rejection region in test (iii). Table 1 reports the size and power functions for tests (i), (ii) and (iii) for all permutations of H and λ drawn from the following sets of values: H={0.5,0.54,0.58,0.62}, λ={1,1.04,1.08,1.12}. The power functions are computed by applying the estimators of H and λ based on (11) and (12) to simulated series generated in accordance with (1) to (4), with NIID random numbers used to generate $m[\eta_t(k)]$ and $u_t$.

Test (i) is oversized if λ>1. If there is multifractality, the test rejects the null hypothesis of no positive or negative persistence too frequently. For H>0.5, the power of test

(i) increases monotonically with H. Test (ii) is undersized if H>0.5. If there is positive persistence and long memory, the test fails to reject the null hypothesis of no multifractality too frequently. For λ>1, the power increases monotonically with λ. Small departures from λ=1 (e.g. λ=1.04) are more easily detected by (ii) than are small departures from H=0.5 (e.g. H=0.54) by (i). The test of (iii) is effective in detecting departures from {H, λ}={0.5, 1}. The power functions increase monotonically with both H and λ. Consistent with (i) and (ii), small departures from λ=1 are more easily detected than small departures from H=0.5.

## 5. Empirical illustration: Fitting the MMAR to foreign exchange rate returns

Table 2 and Figure 2 report results for the estimation and hypothesis test procedures described in Sections 3 and 4, applied to T=5,000 daily logarithmic returns for 12 national currency exchange rates against the US dollar, for the period January 1993 to February 2012. The data were sourced from *Datastream*.

The left-hand panel of Table 2 compares the realized estimates of H and λ based on the log returns series pre-filtered to eliminate short-range dependence by fitting an autoregressive model to the observed returns, with critical values from NIID Monte Carlo simulations. Pre-filtering, however, can produce distortions in the test for long-range dependence. Accordingly, the right-hand panel of Table 2 compares realized estimates based on the unfiltered log returns series with critical values from recursive Monte Carlo simulations. For the latter, we construct 5,000 simulated series using $r^*_{i,t} = \sum_k \hat{\rho}_k r^*_{i,t-k}$ where $u_{i,t}$ are NIID and $\hat{\rho}_k$ are the coefficients from the autoregressive model fitted to the observed returns. Only those values of k in the range 1≤k≤12 for which $\hat{\rho}_k$ is significant at the 0.05 level are included in the construction of $r^*_{i,t}$.

According to Table 2, test (i) fails to reject $H_0$:$H=0.5$ in favour of $H_1$:$H \neq 0.5$ at the 0.05 level for any of the 12 exchange rate log returns series, using either the filtered or the unfiltered returns. By contrast, test (ii) rejects $H_0$:$\lambda=1$ in favour of $H_1$:$\lambda>1$ for 11 of the 12 series, the sole exception being the GB pound. Consistent with the results of (i) and (ii), test (iii) rejects $H_0$:$\{\lambda=1, H=0.5\}$ at the 0.05 level for 11 of the 12 series, the sole exception being the GB pound as before.[1,2] The results for the GB pound, traded in large volume in a highly liquid market, might suggest a connection between multifractality and market efficiency [12]. By contrast, the high estimated $\lambda$ for the Japanese yen, also traded in large volumes but in a market subject to regular intervention by the Bank of Japan [16], might reflect a tendency for central bank intervention to induce jumps in exchange rate volatility [17].

The empirical tests provide no support for the hypothesis that exchange rate returns are characterized by either long memory and positive persistence, or negative persistence. With the exception of the GB pound, however, there is consistent evidence of multifractality of a form consistent with a data generating process of the form $\Delta^{(1)}p_t = u_t$, with $u_t \sim N(0,\Delta\theta(t))$ and $\theta(t)$ defined as in (2). The MMAR, constructed by compounding fractional brownian motion (FBM) with a multifractal time-deformation process, receives qualified support: it appears that exchange rate returns are best described by applying the multifractal time-deformation process to NIID returns.

---

[1] During the 2007-09 financial crisis, the GB pound fell sharply in value against the US dollar. To investigate whether this may have influenced the estimation of H and $\lambda$ for the UK, we repeat the estimations for the sub-period 16/05/1997-01/01/2007. The estimated $\lambda$ is insignificant for both sub-periods, and the result for the GB pound reported in Table 2 does not appear to be related to the financial crisis.

[2] The estimation procedure described in Section 3 is based upon an assumption that $u_t$ are normally distributed. As an informal check on the validity of this assumption, for each of the 12 series we compare the sample kurtosis coefficient with the upper limit of a 95% one-sided confidence interval for the sample kurtosis, obtained by running 5,000 simulations of $\Delta^{(1)}p_t = u_t$ where $u_t \sim N(0,\Delta\theta(t))$ and $\Delta\theta(t)$ is defined in accordance with (2) for $\ln m[\eta_t(k)] \sim N(\lambda,2(\lambda-1)/\ln 2)$. The parameter $\lambda$ used in the simulations is the estimated $\lambda$ for each series, with the sole exception of the UK for which the simulation is based on $\lambda=1$. For all 11 series for which the estimated $\lambda$ in Table 2 is significantly greater than one the sample kurtosis lies within the confidence interval, suggesting that the normality assumption in respect of $u_t$ is reasonable. For the UK, the sample kurtosis of 6.38 lies outside the confidence interval, which suggests that the normality assumption is not appropriate. IID returns drawn from a student-t distribution with approximately six degrees of freedom would replicate the sample kurtosis in the observed UK series.

## 6. Conclusion

In this paper, we develop statistical testing criteria, based on conventional hypothesis testing methodology, to facilitate comparisons between processes that are described by the MMAR, and processes that are characterized by simpler models nested within the MMAR. Monte Carlo simulations are used to generate critical values for the relevant tests. The methods are illustrated by means of an analysis of the multifractal properties of the daily log returns for the exchange rates of 12 currencies against the US dollar for the period 1993-2012. The analysis suggests that 11 of the 12 exchange rate returns series should be described by compounding NIID returns with a multifractal time-deformation process. The GB pound/US dollar exchange rate presents no evidence of multifractality, and future research on this topic is warranted.

Table 1       Size and power functions

| | | Significance level = 0.10 | | | | Significance level = 0.05 | | | | Significance level = 0.01 | | | |
|---|---|---|---|---|---|---|---|---|---|---|---|---|---|
| | H→ | 0.5 | 0.54 | 0.58 | 0.62 | 0.5 | 0.54 | 0.58 | 0.62 | 0.5 | 0.54 | 0.58 | 0.62 |
| λ↓ | | | | | | | | | | | | | |
| (i) Test H$_0$: H=0.5 against H$_1$: H≠0.5 | | | | | | | | | | | | | |
| 1.00 | T=2500 | .100 | .255 | .583 | .836 | .050 | .173 | .473 | .773 | .010 | .067 | .282 | .617 |
| 1.04 | | .167 | .273 | .540 | .789 | .104 | .194 | .443 | .726 | .036 | .089 | .276 | .567 |
| 1.08 | | .225 | .292 | .509 | .738 | .154 | .215 | .423 | .676 | .071 | .106 | .270 | .522 |
| 1.12 | | .267 | .309 | .489 | .693 | .194 | .235 | .407 | .629 | .096 | .124 | .259 | .483 |
| 1.00 | T=5000 | .100 | .317 | .700 | .930 | .050 | .228 | .615 | .893 | .010 | .088 | .395 | .758 |
| 1.04 | | .160 | .317 | .646 | .893 | .101 | .237 | .566 | .843 | .031 | .103 | .364 | .704 |
| 1.08 | | .223 | .327 | .600 | .841 | .154 | .247 | .519 | .787 | .064 | .118 | .337 | .647 |
| 1.12 | | .274 | .340 | .556 | .791 | .203 | .263 | .475 | .735 | .097 | .131 | .315 | .594 |
| (ii) Test H$_0$: λ=1 against H$_1$: λ>1 | | | | | | | | | | | | | |
| 1.00 | T=2500 | .100 | .054 | .027 | .011 | .050 | .024 | .009 | .003 | .010 | .004 | .001 | .000 |
| 1.04 | | .688 | .614 | .527 | .437 | .606 | .515 | .418 | .316 | .443 | .337 | .237 | .152 |
| 1.08 | | .862 | .823 | .773 | .717 | .820 | .767 | .705 | .621 | .715 | .633 | .531 | .426 |
| 1.12 | | .914 | .901 | .879 | .848 | .893 | .868 | .835 | .788 | .823 | .777 | .716 | .628 |
| 1.00 | T=5000 | .100 | .049 | .024 | .005 | .050 | .021 | .005 | .001 | .010 | .001 | .000 | .000 |
| 1.04 | | .791 | .727 | .645 | .558 | .733 | .651 | .556 | .455 | .605 | .495 | .381 | .268 |
| 1.08 | | .931 | .912 | .881 | .839 | .912 | .881 | .842 | .789 | .861 | .802 | .731 | .640 |
| 1.12 | | .969 | .960 | .950 | .930 | .960 | .950 | .929 | .903 | .934 | .909 | .874 | .829 |
| (iii) Test H$_0$: {H=0.5, λ=1} | | | | | | | | | | | | | |
| 1.00 | T=2500 | .100 | .186 | .461 | .759 | .050 | .099 | .326 | .649 | .010 | .018 | .110 | .362 |
| 1.04 | | .581 | .625 | .748 | .883 | .478 | .493 | .620 | .802 | .292 | .241 | .327 | .532 |
| 1.08 | | .826 | .849 | .903 | .954 | .762 | .774 | .831 | .912 | .587 | .567 | .609 | .732 |
| 1.12 | | .919 | .934 | .960 | .980 | .882 | .892 | .925 | .961 | .770 | .756 | .788 | .855 |
| 1.00 | T=5000 | .100 | .235 | .595 | .877 | .050 | .145 | .472 | .809 | .010 | .034 | .232 | .603 |
| 1.04 | | .702 | .759 | .884 | .968 | .617 | .655 | .803 | .931 | .435 | .413 | .564 | .792 |
| 1.08 | | .919 | .942 | .973 | .992 | .883 | .906 | .948 | .984 | .778 | .783 | .850 | .932 |
| 1.12 | | .973 | .978 | .993 | .997 | .959 | .966 | .982 | .996 | .912 | .918 | .947 | .976 |

Table 2      Empirical results

|  | Filtered returns, critical values from NIID simulations | | | | | Unfiltered returns, critical values from AR simulations | | | | |
|---|---|---|---|---|---|---|---|---|---|---|
|  | $\hat{H}$ | $\hat{\lambda}$ | $\pi_H$ | $\pi_\lambda$ | $\pi_{H,\lambda}$ | $\hat{H}$ | $\hat{\lambda}$ | $\pi_H$ | $\pi_\lambda$ | $\pi_{H,\lambda}$ |
| Australia | 0.527 | 1.121 | 0.417 | 0.000 | 0.000 | 0.523 | 1.121 | 0.394 | 0.000 | 0.000 |
| Canada | 0.518 | 1.109 | 0.577 | 0.000 | 0.000 | 0.493 | 1.116 | 0.560 | 0.000 | 0.000 |
| Denmark | 0.510 | 1.076 | 0.730 | 0.000 | 0.004 | 0.509 | 1.076 | 0.742 | 0.003 | 0.010 |
| Israel | 0.496 | 1.130 | 0.980 | 0.000 | 0.000 | 0.515 | 1.122 | 0.840 | 0.000 | 0.000 |
| Japan | 0.495 | 1.182 | 0.968 | 0.000 | 0.000 | 0.509 | 1.175 | 0.851 | 0.000 | 0.000 |
| Norway | 0.495 | 1.071 | 0.974 | 0.001 | 0.013 | 0.486 | 1.073 | 0.970 | 0.004 | 0.025 |
| New Zealand | 0.555 | 1.062 | 0.109 | 0.005 | 0.005 | 0.549 | 1.075 | 0.092 | 0.003 | 0.003 |
| Singapore | 0.514 | 1.116 | 0.644 | 0.000 | 0.000 | 0.534 | 1.130 | 0.469 | 0.000 | 0.000 |
| Sweden | 0.502 | 1.064 | 0.878 | 0.003 | 0.029 | 0.503 | 1.065 | 0.905 | 0.007 | 0.036 |
| Switzerland | 0.478 | 1.115 | 0.613 | 0.000 | 0.000 | 0.487 | 1.117 | 0.810 | 0.000 | 0.000 |
| Taiwan | 0.527 | 1.141 | 0.418 | 0.000 | 0.000 | 0.560 | 1.113 | 0.303 | 0.000 | 0.000 |
| UK | 0.510 | 0.980 | 0.720 | 0.955 | 0.830 | 0.491 | 0.978 | 0.708 | 0.967 | 0.811 |

Graphs for Figure 1: Multifractal spectrum.

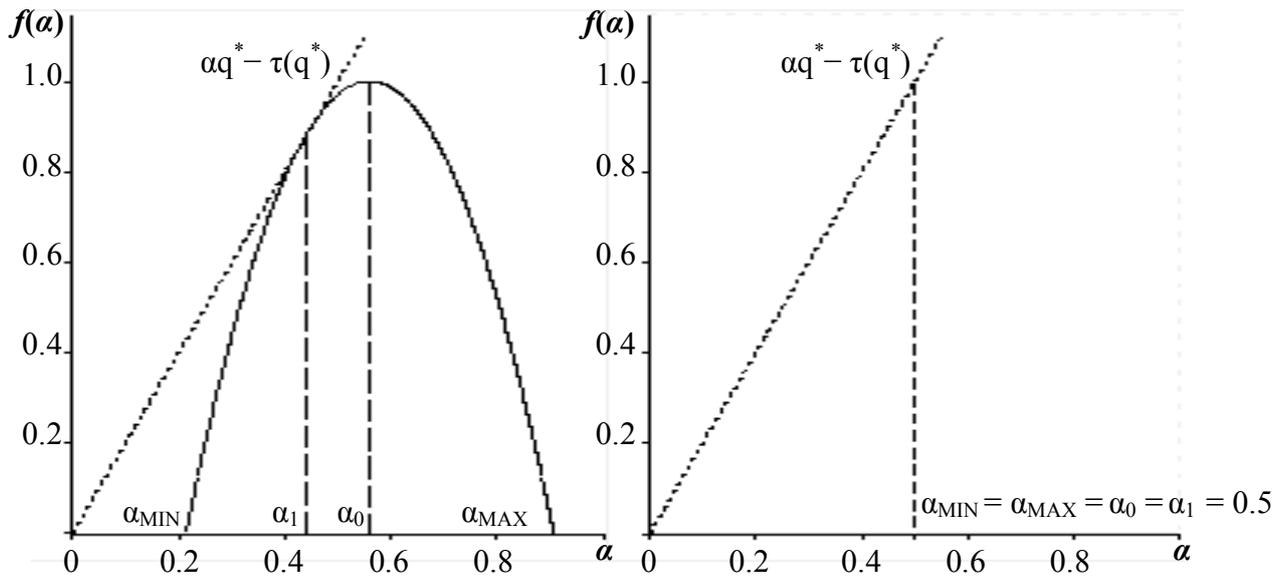

Note

Panel (a) illustrates the multifractal spectrum $f(\alpha)$ for $H=0.5$, $\lambda=1.12$. These values imply $\alpha_0=0.56$, $\alpha_1=0.44$, $\alpha_{MIN}=0.214$, $\alpha_{MAX}=0.906$. Panel (b) illustrates the multifractal spectrum $f(\alpha)$ for $H=0.5$, $\lambda=1$ (the multifractal spectrum degenerates to a point). These values imply $\alpha_0=\alpha_1=\alpha_{MIN}=\alpha_{MAX}=0.5$.

Graph for Figure 2: 95% confidence 'ellipse' (one-sided test for λ; two-sided test for H). Filtered returns, critical values from NIID simulations

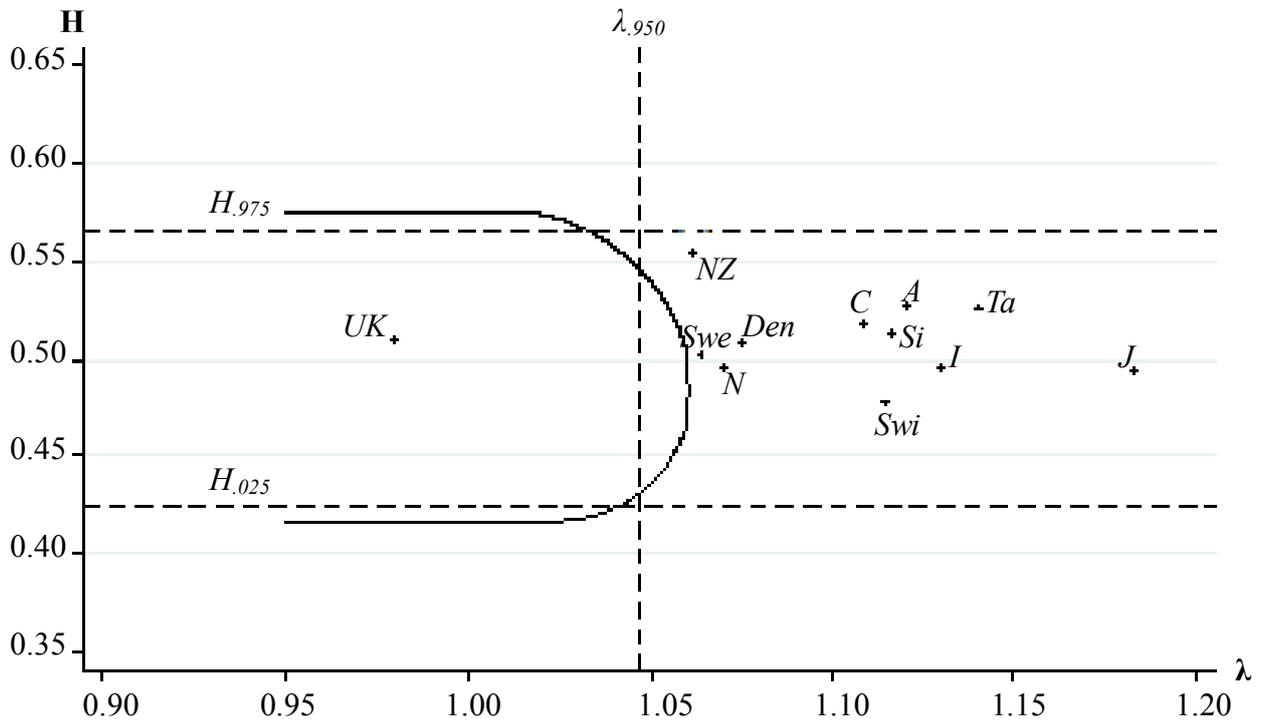

Note

A = Australian Dollar; C = Canadian Dollar; D = Danish Krone; I = Israeli New Sheqel; J = Japanese Yen; N = Norwegian Krone; NZ = New Zealand Dollar; Si = Singapore Dollar; Swe = Swedish Krona; Swi = Swiss Franc; Ta = Taiwanese Dollar; UK = British Pound.

$\lambda_{.950}$ = 5% upper critical value for λ (one-sided test).
$H_{.975}$ = 5% upper critical value for H (two-sided test).
$H_{.025}$ = 5% lower critical value for H (two-sided test).